\documentclass{optica-article}
\journal{opticajournal}

\articletype{Research Article}

\usepackage{subfiles}
\usepackage[section]{placeins}
\usepackage{lineno}
\usepackage[version=4]{mhchem}
\usepackage{siunitx}
\mathchardef\mhyphen="2D
\definecolor{oceanboatblue}{rgb}{0.0, 0.47, 0.75}

\begin{document}

\title{Heterogeneous integration of single InAs/InP quantum dots with the SOI chip using direct bonding}

\author{Marek Burakowski,\authormark{1} Paweł Holewa,\authormark{1,2,3,*}  Aurimas Sakanas,\authormark{2} Anna Musiał,\authormark{1} Grzegorz Sęk,\authormark{1} Paweł Mrowiński,\authormark{1,*} Kresten Yvind,\authormark{2,3} Elizaveta Semenova,\authormark{2,3} and Marcin Syperek\authormark{1}}

\address{\authormark{1}Department of Experimental Physics, Faculty of Fundamental Problems of Technology, Wrocław University of Science and Technology, Wybrzeże Wyspiańskiego 27, 50-370 Wrocław, Poland\\
\authormark{2}DTU Electro, Technical University of Denmark, 2800 Kongens Lyngby, Denmark\\
\authormark{3}NanoPhoton-Center for Nanophotonics, Technical University of Denmark, 2800 Kongens Lyngby, Denmark}

\email{\authormark{*}pawel.holewa@pwr.edu.pl, pawel.mrowinski@pwr.edu.pl} 

\begin{abstract*} 
Quantum information processing with photons in small-footprint and highly integrated silicon-based photonic chips requires incorporating non-classical light sources. In this respect, self-assembled III-V semiconductor quantum dots (QDs) are an attractive solution, however, they must be combined with the silicon platform. Here, by utilizing the large-area direct bonding technique, we demonstrate the hybridization of InP and SOI chips, which allows for coupling single photons to the SOI chip interior, offering cost-effective scalability in setting up a multi-source environment for quantum photonic chips. We fabricate devices consisting of self-assembled InAs QDs embedded in the tapered InP waveguide (WG) positioned over the SOI-defined Si WG. Focusing on devices generating light in the telecom C-band compatible with the low-loss optical fiber networks, we demonstrate the light coupling between InP and SOI platforms by observing photons outcoupled at the InP-made circular Bragg grating outcoupler fabricated at the end of an \SI{80}{\micro\meter}-long Si WG, and at the cleaved edge of the Si WG. Finally, for a device with suppressed multi-photon generation events exhibiting \SI{80}{\percent} single photon generation purity, we measure the photon number outcoupled at the cleaved facet of the Si WG. We estimate the directional on-chip photon coupling between the source and the Si WG to \SI{5.1}{\percent}.
\end{abstract*}

\section{Introduction}
Information processing at the quantum level, based on single photons or electrons, is a fundamental challenge to enable advanced quantum computation. Integrated quantum photonic circuits (IQPCs) offer scalability while operating on multiple flying qubits, simultaneously providing small-footprint devices with unprecedented performance. It can be achieved with complementary metal-oxide-semiconductor (CMOS) technology on large-size silicon-on-insulators (SOI) wafers. Although the SOI platform has already demonstrated scalability and flexibility in integrating and configuring a variety of active and passive photonic devices, including couplers, splitters, combiners, outcouplers, phase shifters and modulators \cite{Dong2014,Xu2022}, its main drawback is the lack of material-compatible light sources emitting non-classical photon. The main reason is silicon's indirect bandgap, preventing efficient light generation, and the significant lattice constant and the thermal expansion coefficient mismatch between Si and III–V materials, which are major challenges for monolithic integration \cite{Komljenovic2018}. The solution is hybridising the SOI platform with  other material system with high-performance non-classical light sources.

Such a hybrid approach is not new, and it is often proposed for the heterogeneous integration of III-V lasers to silicon and realized typically by wafer bonding or transfer printing methods\cite{Hu2019,Yang2012,Liang2010,Jung2016}. In the context of IQPCs, a remarkable effort is paid to including an isolated quantum emitter acting as a source of triggered single photons\cite{Unsleber2016,Ding2016,He2013}, or to control electron spins in crystal adatoms acting  as quantum memories\cite{Loss1998,Elzerman2004,Kroutvar2004},  which are the main components for the realization of fast and efficient quantum algorithms. For efficient photon guiding through Si waveguides, one should consider the issue of the Si absorption. It limits the usable photons' wavelength of above \SI{1.2}{\micro\meter}. Among various single-photon sources emitting $\SI{>1.2}{\micro\meter}$, one can consider erbium ions \cite{Gritsch2023}, defects in 2D MoTe\textsubscript{2} \cite{Zhao2021}, carbon nanotubes \cite{He2018}, or semiconductor quantum dots (QDs) \cite{Arakawa2020}.

Among others, the self-assembled semiconductor QDs are near-ideal sources of single photon states characterized by the high purity of emission\cite{Schweickert2018, Miyazawa2016, Hanschke2018}, and nearly perfect photon indistinguishability\cite{Somaschi2016,Wei2014,Thoma2016}. Furthermore, QDs benefit from the broad wavelength tunability due to bandgap variation and size effect, as well as the capability of generation maximally entangled photon pairs with high fidelity using the biexciton–exciton radiative cascade\cite{Zeuner2021, Mller2018,Moreau2001,Stevenson2006}. Such photon pairs have the ideal two-photon quantum state for advanced schemes of quantum information processing\cite{Ekert1991,Briegel1998,Gisin2007,Simon2007,Mower2013}. Moreover, unlike highly attenuated lasers\cite{Kim2021, Guo2016}, parametric down-conversion \cite{Kwiat1995} or four-wave mixing in nonlinear medium \cite{Yan2015} widely used for quantum information processing, QDs can generate non-classical states deterministically\cite{Arakawa2020}. 

The application potential of III-V QDs for on-chip photonics was revealed by several recent demonstrations also on its hybrid integration with silicon nitride (\ce{Si3N4}) platform, which includes highly efficient optical interface between heterogeneously integrated \ce{Si3N4} waveguides (WGs) and GaAs-based QDs\cite{Davanco2017,Schnauber2019}. Another interesting example is a single QD in a nanowire transferred by a nanomanipulator and integrated with the \ce{Si3N4}-based photonic circuit\cite{Zadeh2016}. Next, epitaxially grown InP-based QDs emitting in the O-band telecom window (\SI{1.3}{\micro\meter} wavelength) were positioned deterministically on a silicon photonic chip with nanoscale precision by a pick-and-place approach\cite{Kim2017}, as well as an InAsP QD emitting around \SI{0.9}{\micro\meter} was integrated in a CMOS compatible SiN photonic circuit with reconfigurable on-chip single-photon filtering and wavelength division multiplexing\cite{Elshaari2017}. However, demonstrations with the heterogeneous integration of III-V QD-based single photon source with the SOI platform are very limited~\cite{Katsumi2022_springer}. 

In this work, we present the fabrication and characterization of an optically-triggered on~chip single-photon source made of an InAs/InP QD embedded in the InP nanobeam WG heterogeneously integrated with the Si WG on an SOI chip. The Si WG ends with an outcoupler based on half of the circular Bragg grating (CBG), allowing the outcoupling of photons to an off-chip environment. Therefore, the device comprises three essential elements of a IQPC: the SPS, the low-loss photons guiding medium, and the outcoupler. To hybridize the InP nanobeam with the SOI chip, we successfully developed the technological process involving QDs growth, photo- and electron-beam-lithography, dry and wet etching, and InP to SOI direct wafer bonding. Importantly, we utilise the advantage of InAs/InP QDs naturally emitting photons at a technologically-important telecom C-band, providing extremely low losses in Si WGs as well as in silica fibers. We characterize the device optically to demonstrate the effective transfer of photons generated by the biexciton-exciton cascade in the QD to the Si WG. We measure the single photon purity of the source, by outcoupling photons off the chip, and the photon coupling efficiency between the source and the Si WG. The latter is compared to the results of calculations based on the finite-difference time-domain (FDTD) simulations. Our results pave the way for a new generation of quantum devices combining optically active heterogeneously integrated InP/Si waveguides containing a single InAs/InP QD and a passive circuit realized on the Si platform with additional elements like ring resonators, directional couplers for efficient on-chip routing of light quanta, single photon detectors\cite{Kahl2015} or grating-based outcouplers for interconnection with external fiber network.

\begin{figure}[!htb]
\centering\includegraphics[width=0.7\textwidth]{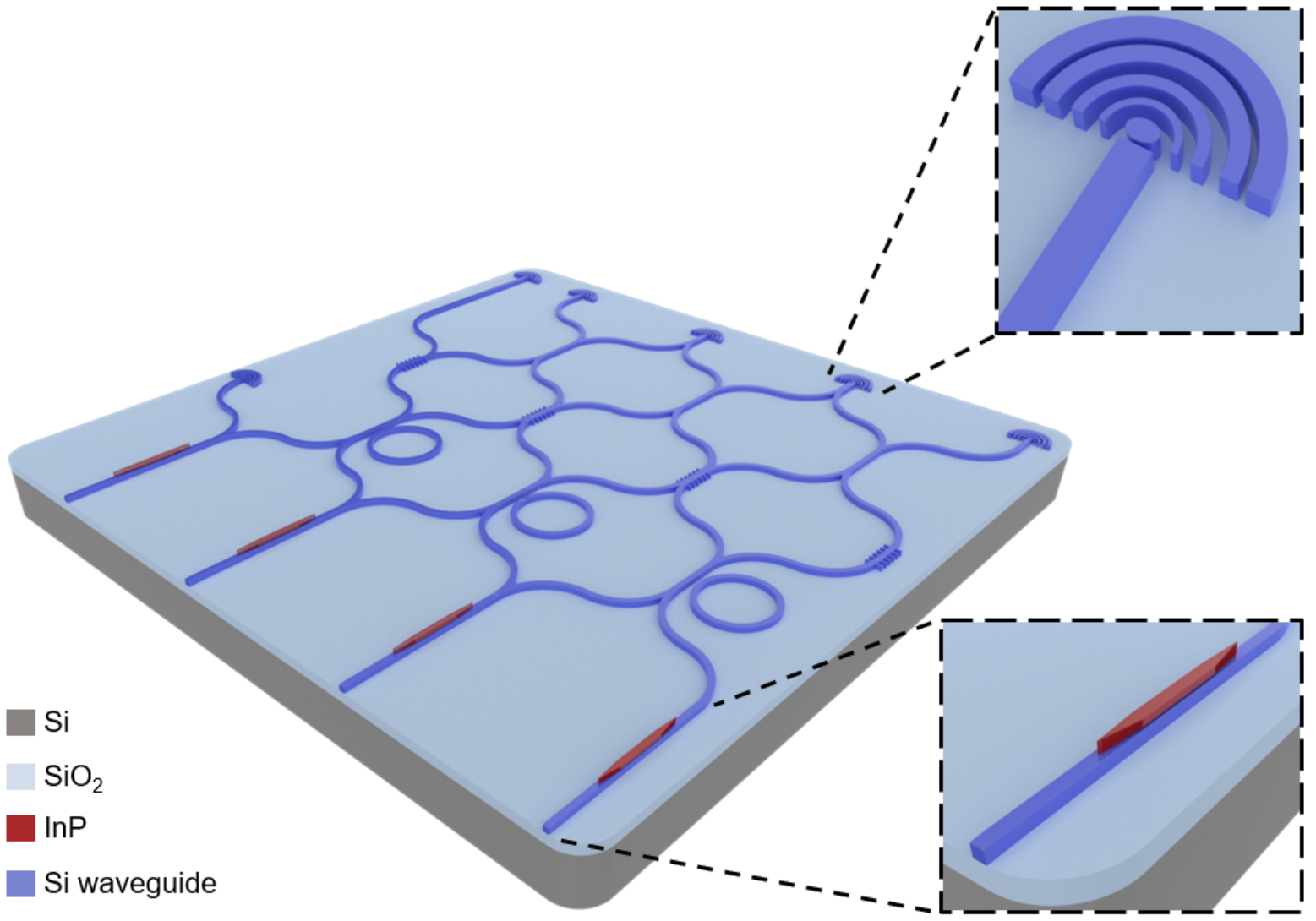}
\caption{Conceptual illustration of the integrated quantum photonic circuit (IQPC) with heterogeneously integrated QD emitter. The bottom layer (gray) is silicon substrate, the center layer (light blue) is silica, and the top layer consists of patterned silicon waveguides (blue) and InP (red) with InAs QDs. Enlarged images demonstrate a single photon source and outcoupler regions, while in between a circuit based on silicon waveguides presents a possible realization of information processing based on phase shifters, ring resonators and directional couplers.}
\label{fig:Figure1}
\end{figure}

\section{Device structure}
The investigated device contains three important elements of a IQPC embedded in the SOI platform: (i) the SPS, (ii) the low-loss photon guiding medium, and (ii) the photons outcoupler off the chip. Figure~\ref{fig:Figure1} presents the conceptual scheme of an exemplary IQPC designed to deliver processed signals off the chip to free space or to silica fibres network operating in the telecom C-band. Within the scheme, the SPS must be positioned accurately over the Si WG to enable efficient coupling of emitted photons into on-chip photonic circuits. Subsequently, photons are guided by the low-loss Si WGs and redistributed over various on-chip passive and active elements, allowing for signal processing on a single photon level. Finally, the processed states are transfered off the chip by light scattering on the outcoupler, read out by off-chip photon detectors, or transmitted through the silica fibres over the quantum network.

\begin{figure}[!htb]
\centering\includegraphics[width=\textwidth]{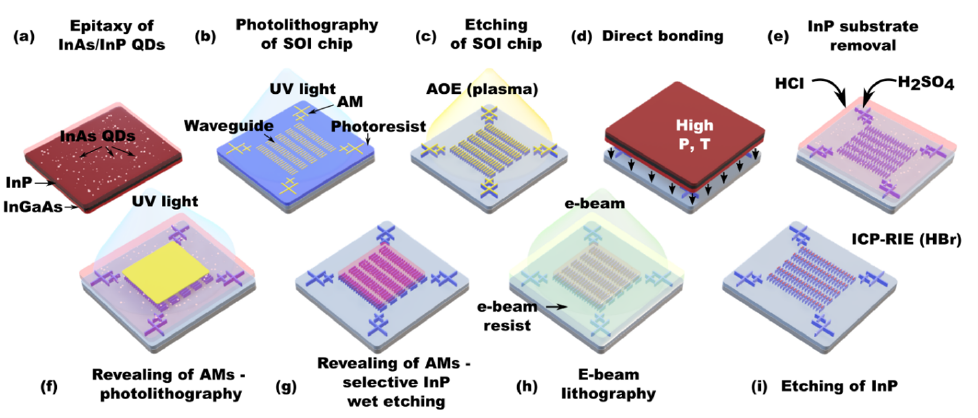}
\caption{Fabrication workflow of the hybrid InP/Si waveguides based on the scalable molecular wafer bonding technique. (a) Epitaxy of low-density InAs/InP QDs. (b) Patterning of Si waveguides and alignment marks (AM) in UV lithography. (c) Pattern transfer to Si. (d) Direct bonding of the SOI and InP wafers. (e) Substrate removal in wet etching. (f-g) Additional UV lithography and wet etching steps to reveal AMs in SOI, (h) E-beam lithography of the InP tapers (i) pattern transfer to InP using ICP-RIE dry etching and subsequent cleaving of the wafer perpendicularly through the Si waveguides.}
\label{fig:Figure2}
\end{figure}

The essential step of the proposed device fabrication technology is wafer bonding of the InP and SOI platforms. The technology facilitates the light transfer from the InP to the Si waveguide and is a reliable solution for the cost-effective fabrication of IQPCs on a large scale. We have outlined the technological steps in Fig.~\ref{fig:Figure2}. The process starts with the growth of low-density Stranski-Krastanov InAs/InP QDs using metal-organic vapour phase epitaxy (MOVPE)\cite{berdnikov2023fine} (Fig.~\ref{fig:Figure2}a). The QDs are placed in the middle of a 580-nm-thick InP slab, which serves as a platform for defining InP WGs. An InGaAs etch stop layer separates the slab from the InP substrate. Concurrently, Si WGs and alignment marks (AMs) are patterned using ultra-violet (UV) optical lithography in a positive-tone photoresist on top of the 220-nm-thick Si layer of a standard SOI chip (Fig.~\ref{fig:Figure2}b). The Si WGs are defined by two \SI{10}{\micro\meter}-wide openings which also act as efficient vertical out-gassing channels to facilitate high area bonding\cite{Liang2008}. The pattern is transferred using inductively coupled plasma advanced oxide etch (ICP - AOE) to the Si layer (Fig.~\ref{fig:Figure2}c). Then, an oxygen plasma treatment of the flipped InP chip and Si surfaces is used to increase the density of hydroxyl groups, which are responsible for establishing bonds at the interface\cite{Pll1999,Tong1996}. Direct contact between surfaces is established immediately after the low-power plasma ashing by high force ($\SI{\sim2}{\kilo\newton}$) and temperature (\SI{250}{\celsius}) in vacuum conditions in the bonder. This effectively creates stable contact between InP and Si without any intermediate layer (Fig.~\ref{fig:Figure2}d), as evidenced by a typically high bonding yield of more than \SI{70}{\percent}, evaluated for a quarter of a 2-inch InP wafer.

The InP substrate and etch stop layers are removed using HCl and \ce{H2SO4$:$H2O2}, respectively (Fig.~\ref{fig:Figure2}e). In order to achieve high-accuracy positioning of InP nanobeam WGs on top of the Si WGs during an e-beam lithography step, additional processing steps are introduced. First, the AMs defined in SOI are revealed. Openings in InP are defined using UV lithography (Fig.~\ref{fig:Figure2}f) and then transferred by HCl etching (Fig.~\ref{fig:Figure2}g). E-beam lithography using hydrogen silsesquioxane (HSQ) negative tone resist with high accuracy alignment (approx. \SI{40}{\nano\meter})\cite{Sakanas2019} is used to define the InP nanobeam WGs and outcouplers positioned on top of the Si waveguides (Fig.~\ref{fig:Figure2}h). This pattern is transferred to InP using HBr plasma in ICP-reactive ion etching (RIE) and the HSQ mask is removed in buffered hydrogen fluoride (BHF) (Fig.~\ref{fig:Figure2}i).

Within this approach, we realized a series of hybrid InP/Si WG structures with different geometrical parameters based on the optimized heterogeneously integrated InP-Si on-chip system\cite{Mrowiski2023}, enabling broadband on-chip directional light coupling from the QD to the Si WG of approx.~\SI{33}{\percent} efficiency in the spectral range of $\SIrange{1.5}{1.6}{\micro\meter}$. The assumed geometry of crucial elements in the IQPC design is a \SI{0.58}{\micro\meter}-high InP WG on top of a \SI{0.22}{\micro\meter}-high Si WG. The InP(Si) WG width is set to 1.5(2.0) \textmu m, 1.5(3.0) \textmu m, 2.0(3.0) \textmu m, 1.5(4.0) \textmu m, 1.5(5.0) \textmu m, and 3.0(5.0) \textmu m. The straight hybrid WG section is \SI{100}{\micro\meter} long, and we restrict the geometry of the InP WG taper to \SI{25}{\micro\meter} length and finite tip width down to \SI{0.25}{\micro\meter}. The outcoupler is realized with half of the CBG consisting of InP rings on top of a planar Si section (alternatively, a similar system could be realized directly in Si). The grating geometry contains 5 periods in which the ring width is \SI{0.7}{\micro\meter}, and the grating period is \SI{1.2}{\micro\meter}. 

\begin{figure}[!htb]
\centering\includegraphics[width=\textwidth]{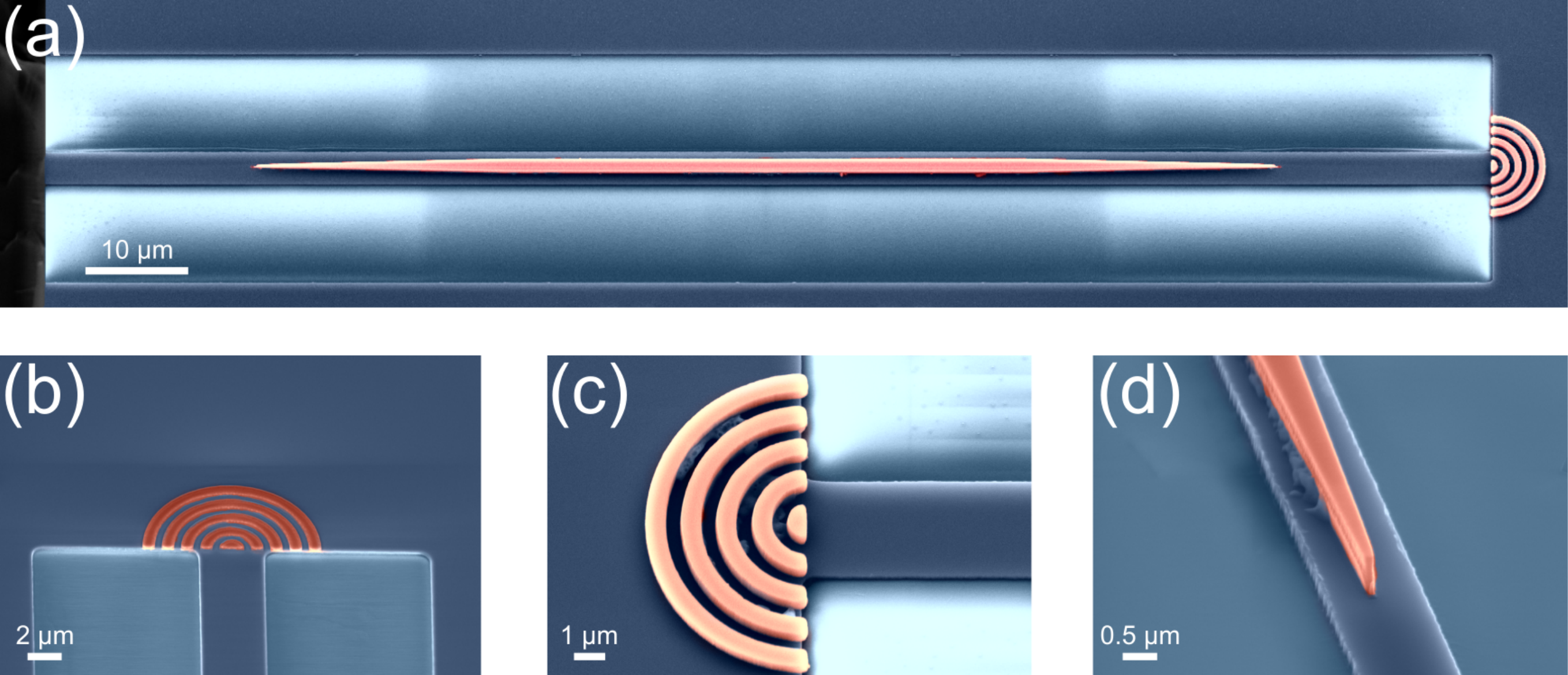}
\caption{(a) Scanning electron micrographs of the hybrid InP/Si waveguide structure  containing a self-assembled InAs/InP QD emitter, (b) and (c) circular Bragg grating (CBG)-based outcoupler, (d) linear taper of the InP WG atop of the Si WG.}
\label{fig:Figure3}
\end{figure}

The fabricated hybrid WG geometry is verified by scanning electron microscopy (SEM) analyses. Exemplary micrographs for the on-chip microstructures are demonstrated in Fig.~\ref{fig:Figure3}. In Fig.~\ref{fig:Figure3}a, we present a single device which contains the hybrid WG section, the Si WG section and half of the CBG outcoupler on the right-hand side. On the left-hand side, the structure ends with a cleaved facet. The CBG outcoupler and the InP WG taper structure are depicted in more detail in Fig. \ref{fig:Figure3}b, c and d, respectively. The images show high quality of the realized device and high accuracy of the integrated InP WG on Si WG with an estimated positioning error of less than 100 nm. However, we observed that the final geometry of the device differs slightly from the design, namely, the InP WG width is smaller by approx. 50 nm, while the Si WG width is smaller by approx. 150 nm (difference due to distinct dry etchings), both evaluated by a median value.

\section{FDTD simulations}
The fabricated series of structures described in the previous section were realized for the geometries according to our prior numerical studies, which were performed via a FDTD solver\cite{Mrowiski2023}. We found that the multimode operation of a hybrid waveguide is inevitable due to refractive indices contrast (n\textsubscript{Si} > n\textsubscript{InP}) in the target spectral range. It leads to a more efficient dipole coupling with the propagating modes on the order of \SI{30}{\percent} with the minimum width of the InP WG of \SI{1.2}{\micro\meter}. We also found that a taper length in the range of $\SIrange{15}{25}{\micro\meter}$ provides optimum transfer efficiency from an InP/Si WG to a Si WG, and a taper tip must be of finite width, e.g. \SI{0.25}{\micro\meter}.

The 3D FDTD calculations were based on the exact geometry of the device presented in Fig.~\ref{fig:Figure3}. We use the same numerical procedure as in ref. \cite{Mrowiski2023}. The model structure consists of a straight hybrid WG system defined by its cross-section with InP WG height of \SI{0.58}{\micro\meter}, Si WG height of \SI{0.22}{\micro\meter} and for all structures, InP(Si) WG widths are now corrected according to the analysis performed on SEM micrographs which reads 1.45(1.85) \textmu m, 1.45(2.85) \textmu m, 1.95(2.85) \textmu m, 1.45(3.85) \textmu m, 1.45(4.85) \textmu m, 2.95(4.85) \textmu m.

In the first part, we focus on modelling the dipole emission coupling into the hybrid InP/Si WG. The dipole is placed in the middle of the InP WG and mimics the InAs QD embedded in InP. The coupling coefficient ($\eta$) is calculated as an overlap between the propagating transverse electric (TE) modes and dipole emission over a distance of \SI{30}{\micro\meter} along the WG to ensure convergence, see Supplementary Information (Supp. Inf.) for details – Sec. 2. The directional dipole coupling coefficient $\left(\frac{\eta}{2}\right)$ is around $\SIrange{30}{40}{\percent}$, as shown in Fig.~S1.

\begin{figure}[!htb]
\centering\includegraphics[width=\textwidth]{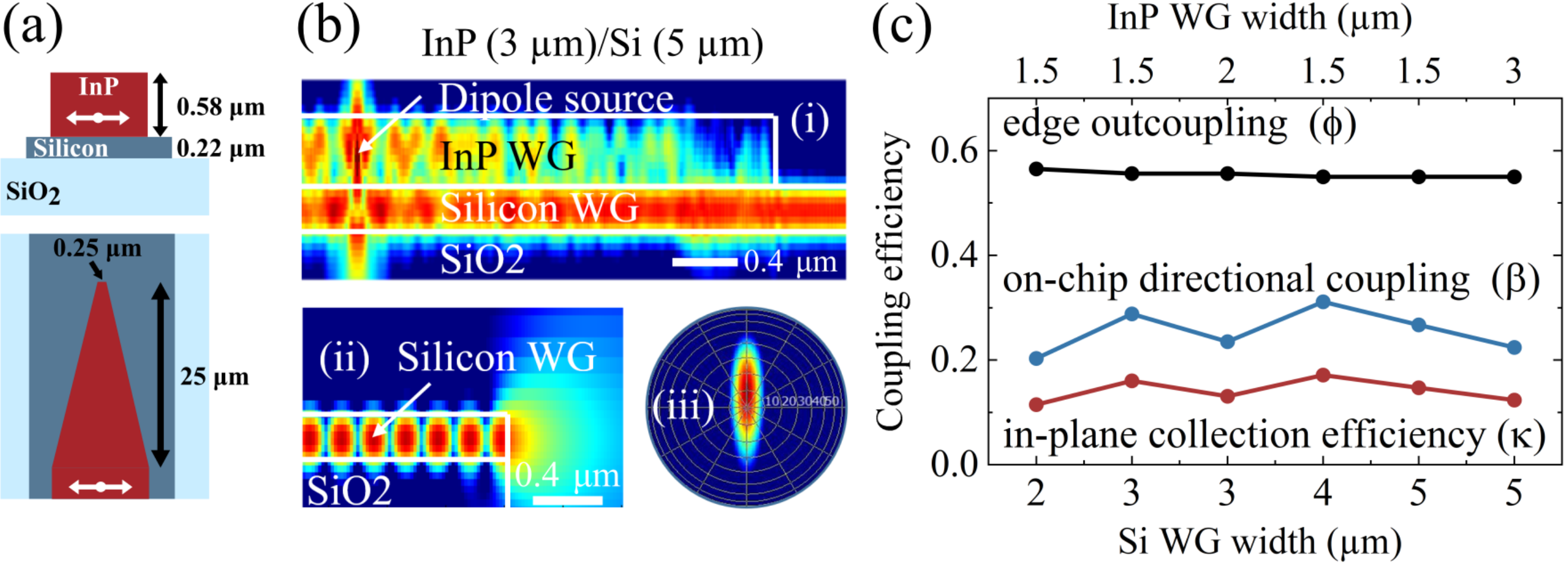}
\caption{a) Schematic model of the hybrid InP/Si waveguide structure on \ce{SiO2} used in FDTD simulations, b) Electric field intensity distribution in a selected InP/Si WG structure used for evaluation of (i) on-chip dipole coupling, (ii) edge outcoupling to the far-field used in micro-photoluminescence (\SI{}{\micro PL}) experiment. c) Edge outcoupling - $\phi$, on-chip directional dipole coupling - $\beta$ and the in-plane collection efficiency - $\kappa$ for all fabricated geometries (with lines as a guide to the eye).}
\label{fig:Figure4}
\end{figure}

Next, we study the same geometries with a taper structure along a InP WG to evaluate the QD emission transfer efficiency between InP/Si WG and Si WG. Here, a straight waveguide region is \SI{4}{\micro\meter}-long and follows a \SI{25}{\micro\meter}-long taper ending with a \SI{0.25}{\micro\meter}-wide tip (see Fig.~\ref{fig:Figure4}a). First, in Fig.~\ref{fig:Figure4}b - panel (i), we demonstrate the cross-sectional $\left | E \right |^{2}$ field distribution of the dipole emission, which indicates an efficient transfer from the hybrid InP/Si WG to Si WG via the taper structure, which allows for the evaluation of the on-chip directional coupling to the Si WG $\left( \beta=\frac{\eta}{2} \times \chi \right)$, where $\chi\sim\SIrange{60}{80}{\percent}$ is taper transmission~\cite{Mrowiski2023}. Second, in Fig.~\ref{fig:Figure4}b - panel (ii), we show $\left | E \right |^{2}$ for the Si WG approaching the cleaved facet, where we used the Si WG structure only to simulate multimode light propagation towards the cleaved facet. Last, in Fig.~\ref{fig:Figure4}b - panel (iii), we demonstrate the evaluation of a far-field distribution of light scattered on the sample edge, which gives insight into the angular distribution of the outcoupled field that might be collected by the microscope objective focused on the waveguide end (it corresponds to the results collected in our experimental configuration, in which we excite from top of the structure and detect from its sidewall). We found that by limiting the collection angle to 45 degrees (provided in the experiment by the 0.65 numerical aperture microscope objective), the evaluated far-field collection efficiency of \SI{89}{\percent} can be reached. However, our simulations show (Fig.~\ref{fig:Figure4}c - black data points) that the edge outcoupling ($\phi$) is on the level of $\SIrange{55}{60}{\percent}$ while the losses are related to the $\SI{\sim30}{\percent}$ reflection of the guided light beam on the interface. The demonstrated evaluation shows that our device allows for the in-plane collection efficiency ($\kappa=\beta \times \phi$) of the directional emission (including losses) on the level of $\SIrange{10}{15}{\percent}$ (Fig.~\ref{fig:Figure4}c - red data points), depending on the geometry. Notably, this value could be compared to the experimental count rate and used to evaluate the efficiency of the on-chip coupling of the QD emission for the fabricated structures. 

\section{Optical experiments}

The optical experiments allow us to validate proposed technological processes and prove the coupling between the QD-based single photon emitter and the SOI platform in our device. Since the QDs density is low, the averaged dot-to-dot separation is \SI{0.42}{\micro\m}, translated on the low-fabrication fidelity defined by the number of devices with a QD. The fabrication fidelity can be increased by using the deterministic fabrication of an InP WG over a localized QD emitter. However, this issue is not addressed in the article. To determine fabrication fidelity, we performed a high-spatially resolved photoluminescence (\SI{}{\micro PL}) experiment in which we tested roughly 50 neighbouring devices to find spectral features related to the presence of a QD in the InP/Si WG. Moreover, we focused on the QDs emitting in the telecom C-band near \SI{1550}{\nano\meter} to hold compatibility with the low-loss silica fiber transmission window. We found that \SI{4}{\percent} of the devices fulfil these requirements. Subsequently, we focused on a single device to examine a QD emission pattern and the QD single photon emission purity and demonstrate the coupling of photons from the InP/Si WG to the Si WG in the SOI platform.

\begin{figure}[!htb]
\centering\includegraphics[width=\textwidth]{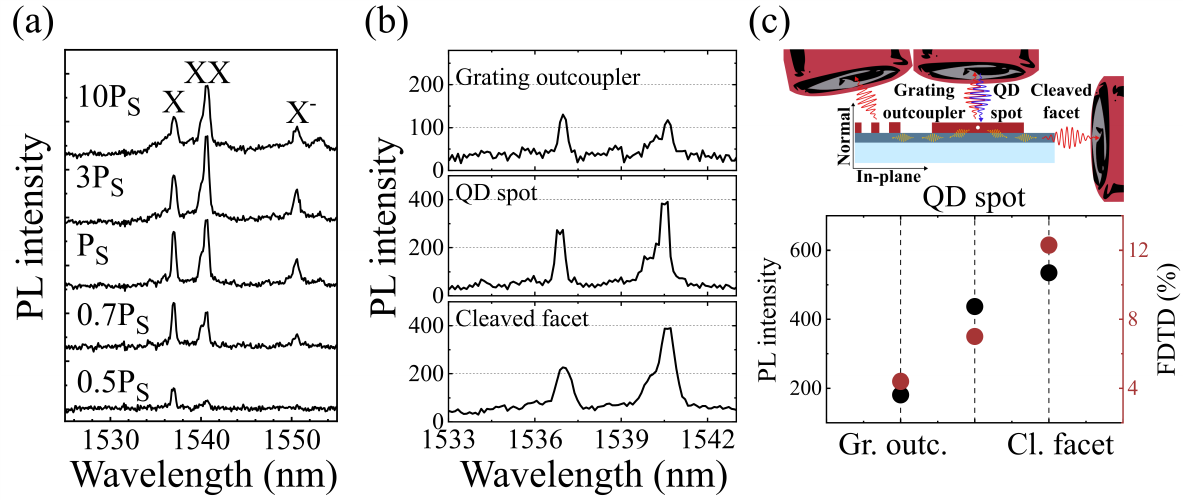}
\caption{(a) \SI{}{\micro PL} spectra at $T=\SI{5}{\kelvin}$ collected under continuous wave excitation versus excitation power on a single quantum dot (QD) integrated into a hybrid InP/Si waveguide structure. The neutral exciton (X), biexciton (XX), and negatively charged exciton (X$^-$) lines are observed at $P_{S} = \SI{7}{\micro\watt}$. (b) QD emission is collected in three different setup configurations, showing qualitatively the same response with clear separation of the excitation and detection path using either the CBG-based outcoupler or cleaved facet of the chip. c) Qualitative comparison of PL intensity collected from the grating outcoupler, right at the QD position and cleaved facet (in-plane direction) with evaluated efficiencies via the FDTD method.}
\label{fig:Figure5}
\end{figure}

Figure~\ref{fig:Figure5}a shows the low-temperature ($T=\SI{5}{\kelvin}$) \SI{}{\micro PL} spectrum from a selected device consisting of a hybrid InP/Si WG system of InP(Si) WG widths of 2.95(4.85)~\SI{}{\micro \m}. The spectrum is registered at different excitation power densities ($P$) at the straight section of the hybrid WG and unveils inhomogeneously broadened emission lines within the $\SIrange{1528}{1553}{\nano\meter}$ spectra range. Based on the previous studies on InAs/InP QDs~\cite{Holewa2022,Holewa2022nano,Zieliński_2013} and analysis of the intensity dependence between the lines, the spectral features are attributed to recombination of the neutral exciton (X), biexciton (XX) and negatively-charged exciton (X$^-$) complexes confined to a single InAs QD placed in the InP WG. In this case, the \SI{}{\micro PL} experiment configuration utilises the same excitation and detection path parallel to the WG surface (normal direction). Importantly, except for the previously mentioned case, the \SI{}{\micro PL} response did not reveal any additional spectral lines in the C-band upon scanning the excitation spot along the WG. This indicates no other optically active QDs emitting in the required spectral range within the device.

To demonstrate the coupling of photons from the QD to the SOI platform, we compared the QD spectra obtained under the already introduced normal \SI{}{\micro PL} excitation-detection scheme with the spectra recorded with two new experimental configurations, where the excitation path is spatially separated from the detection one. In both new schemes, the excitation is realized normal to the InP WG, however, the detection is either provided through the outcoupler or the cleaved facet of the chip, as presented in Fig.~\ref{fig:Figure5}c. This is feasible by applying another microscope objective in the detection path, gathering photons from the outcoupler in the normal direction and the cleaved facet of the Si WG in the in-plane direction of the chip. 

Figure~\ref{fig:Figure5}b displays the results of the \SI{}{\micro PL} experiment for different detection paths. It shows the same qualitative picture of the X and XX emission lines from the InAs QD in the selected device. The X and XX photons registered at the cleaved facet of the chip may have a component directly related to the anisotropic QD emission collected by a microscope objective oriented in-plain of the structure, which makes the coupling between a QD and the Si WG questionable. However, observing out-scattered photons from the Si WG on the grating outcoupler provides solid proof for the coupling. In this case, the QD-emitted photons must travel roughly \SI{80}{\micro\meter} distance separating InP WG and the CBG to be scattered and collected in the normal direction.

Analysis of the emission line intensity presented in Fig.~\ref{fig:Figure5}b can shed light on the coupling efficiency and be qualitatively compared with the FDTD model predictions. The model shows that, for a dipole source placed in the InP WG, photon extraction efficiency towards the normal direction and within the 0.4 numerical aperture (NA) amounts to \SI{7}{\percent}. It is nearly twice less than for the in-plane of the chip. In this case, the photon extraction efficiency is the highest, reaching \SI{12.3}{\percent}, and results from the multiplication of the evaluated edge outcoupling $\phi=\SI{55}{\percent} $ (Fig. 4c) and the on-chip directional coupling $\beta=\SI{22.4}{\percent}$ \cite{Mrowiski2023}. The lowest expected photon extraction efficiency is delivered by the grating outcoupler and reaches \SI{4.4}{\percent}. The value results again from the on-chip directional coupling $\beta=\SI{22.4}{\percent}$ from the dipole source to the hybrid WG, multiplied with the outcoupling efficiency along sample normal direction within the 0.4 numerical aperture of approx. \SI{20}{\percent}~\cite{Mrowiski2023} . 

The numerically evaluated extraction efficiencies for each detection channel are summarized in Fig.~\ref{fig:Figure5}c, in which we additionally plot the integrated intensities of \SI{}{\micro PL} lines from Fig.~\ref{fig:Figure5}b. Both data sets follow the same trends, demonstrating good qualitative agreement, thus confirming extraction efficiency gradation for different detection schemes.

\begin{figure}[!htb]
\centering\includegraphics[width=0.7\textwidth]{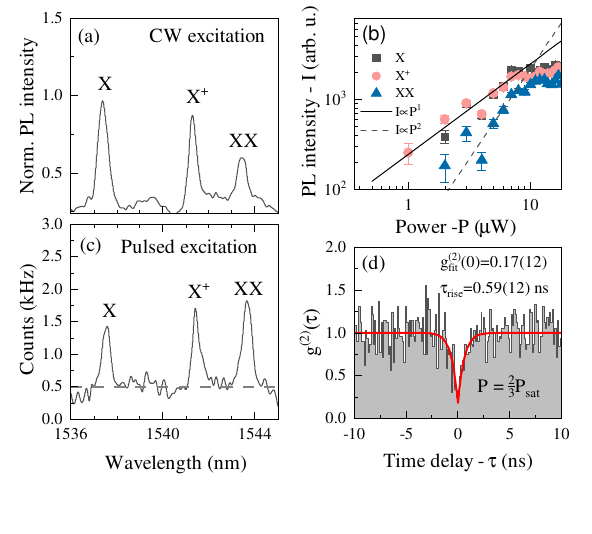}
\caption{(a) \SI{}{\micro PL} spectrum of QD emission collected from a cleaved facet of a device at $T=\SI{5}{\kelvin}$ using continuous wave (CW) excitation, revealing neutral exciton (X), positively-charged exciton (X$^+$), and biexciton (XX) lines. (b) CW excitation power-dependent intensity evolution for the QD emission lines. The solid (dashed) black line shows the linear (superlinear) trend expected for X and X$^+$ (XX) emission from a QD. (c) QD emission under pulsed excitation with the count rate for each QD line. The dashed black line indicates the dark counts' background at \SI{0.5}{\kilo\hertz}. (b) Auto-correlation histogram of the QD natural exciton emission measured under CW excitation and from the cleaved edge of the device. $P_{sat}=\SI{7}{\micro\watt}$. The solid red line is a fit to experimental data.}
\label{fig:Figure6}
\end{figure}

To estimate coupling efficiency quantitatively, a standard continuous-wave \SI{}{\micro PL} is complemented with the pulse excitation experiment, allowing for accurate counting of emitted photons with a single-photon detector (see Supp. Inf.). Experimentally, the in-plane collection efficiency can be found using the following expression $\kappa =\left(N_{phot})/(L_{rep}\times\gamma\right)$, where $N_{phot}$ is the number of collected photons per second (detector counts), $L_{rep}$ is the laser pulse repetition frequency, which equals to \SI{76}{\mega\hertz}, and $\gamma$ is the the photon collection efficiency of the setup. Here, the counted photons were outcoupled from the cleaved facet of the chip after being transmitted through the Si WG and collected by the microscope objective with NA = 0.65. The evaluated $\gamma$ parameter amounts to $\SI{0.115}{\percent}$, as shown in the Supp. Inf (see section 4). For that experiment, another QD-containing device is selected, characterized by the \SI{}{\micro PL} spectrum settled in the telecom C-band as displayed in Fig.~\ref{fig:Figure6}a. The spectrum consists of three inhomogeneously broadened lines assigned to neutral exciton (X), positively-charged exciton (X$^+$) and biexciton emission from the same QD. The assignment is suggested by the expected charged-exciton and biexciton binding energy for InAs/InP QD~\cite{Holewa2022nano,holewa2020prb,Zieliński_2013} and the power-dependent \SI{}{\micro PL} experiment with results presented in Fig.~\ref{fig:Figure6}b. From Fig.~\ref{fig:Figure6}b, one can see that X and X$^+$ emission follow a linear dependence of the line intensity ($I$) with the excitation power ($P$): $I\propto P^1$, whereas for the XX emission, this tendency is close to superlinear: $I\propto P^2$, as expected for the XX-X cascade emission process in a QD~\cite{sek2010,Holewa2022nano,holewa2020prb}. Fig.~\ref{fig:Figure6}c shows the \SI{}{\micro PL} spectrum for the investigated device recorded for the pulsed excitation condition. It allows to establish the count rates above the background level for the X and X$^+$ emission lines, which are $N_{phot}$=\SI{1.0}{\kilo\hertz} and \SI{1.4}{\kilo\hertz}, respectively. Subsequently, it provides the $\kappa=\SI{2.8}{\percent}$. We note here that there can be several factors which lead to underestimating outcoupling efficiency theoretically assessed to \SI{12.3}{\percent} (Fig.~\ref{fig:Figure4}c). Among these factors there are technological ones: i) misalignment of the QD with respect to the InP WG axis\cite{Mrowiski2023}, as the fabrication was not realized in a deterministic manner, which may result in less QD emission coupling to propagating modes in the WG; ii) misalignment between the InP WG and the Si WG, which leads to inefficient on-chip coupling, i.e. transfer of emitted photons from the InP/Si hybrid section to Si WG; iii) irregularities along both InP and Si WGs sections, inaccurate taper sharpening in the vicinity of the taper tip or roughness at the sample edge due to cleaving; iv) lower than expected photon collection efficiency from the cleaved facet (FDTD gives \SI{89}{\percent}); and v) possible non-radiative recombination in the QD which implies reduced internal quantum efficiency~\cite{PhysRevB.77.073303,LodahlStobbe+2013+39+55, Grose2021}.

In this situation, we can estimate the on-chip directional coupling that reaches up to \SI{5.1}{\percent} ( $\beta$ = $\kappa$ /$\phi$ ). However, if the examined QD-WG system is far from optimized, especially considering losses related to before-mentioned factors (i) and (ii), which could reduce $\beta$ by roughly \SI{20}{\percent} \cite{Mrowiski2023}, we could state that further optimization is possible and can lead to more efficient transfer to the Si WG. Noteworthy, since we observe the QD emission from the CBG-based outcoupler, the estimated coupling is only half of the bidirectional emission ($\eta$) coupled to the Si WG. Therefore, we should note that using an appropriately designed reflector (e.g. the Bragg grating) on one side of the WG, providing reflectivity close to \SI{99}{\percent} \cite{Gr2022}, one can assure directional on-chip coupling $\beta$ to a Si WG of approximately \SI{10}{\percent}.

Finally, we investigate the quantum nature of the on-chip coupled QD emission in a selected device by performing second-order auto-correlation measurements in the Hanbury-Brown and Twiss (HBT) configuration. The QD emission is collected from the cleaved facet of the device, while the CW laser excitation is provided from the top at the QD position. We focus on the neutral exciton emission, presented in Fig.~\ref{fig:Figure6}a.

The measurement conditions maximize the signal-to-background ratio and allow for registering the histogram counts, exhibiting a clear antibunching behaviour presented in Fig.~\ref{fig:Figure6}d. The fitting procedure results in the second-order correlation function at zero time delay of $g_{fit}^{(2)}(0)=\SI{0.17(12)}{}$, proving a single photon emission of the source. The observed value of $\tau_{rise}=\SI{0.59(12)}{\nano\second}$ reflects relatively high effective pump rate (laser excitation power)\cite{Holewa2020_SR} and is in accordance with previous results for these QDs\cite{Holewa2020}. In addition, we observe no bunching for the time delays close to $\SI{\pm1}{\nano\second}$, which indicates a low probability of re-excitation of the QD due to carrier recapturing processes in the QD occurring right after the excitonic recombination of carriers trapped by the surrounding traps\cite{Suffczyski2006,Reimer2016}.

\section{Summary and Conclusions}
The InAs/InP quantum dots are perceived as quantum emitters, able to deliver single photons in the telecom C-band spectral range for quantum information processing. In this work, we show the method for successful heterogeneous integration of these single photon emitters with the SOI platform - one of the most interesting for the highly integrated quantum photonic chips. The proposed integration scheme utilizes a wafer-scale direct bonding of III-V optically active materials with a pre-patterned SOI platform, followed by the fabrication of nanobeam waveguides supporting light transfer from an InAs/InP QD to the Si waveguide. This technology paves the way for fast and reliable engineering of quantum photonic chips, providing a step to scalable technology comprising multiple non-classical light sources based on InAs/InP QDs and various passive and active elements fabricated on the SOI platform. 

The performed 3D FDTD numerical calculations allowed setting up a single device geometry comprised of a single InAs QD embedded in the InP tapered waveguide on top of the Si waveguide. The geometry provided on-chip directional coupling efficiency between a QD and the Si waveguide at $\SIrange{20}{30}{\percent}$. This value is reasonable concerning unfavourable refractive index contrast between InP and Si materials. Additionally, with simulations, we pinpointed sources of photon losses providing further device optimization paths, whereas smaller waveguides should achieve better performance.

Finally, we tested selected devices to provide qualitative and quantitative information regarding the on-chip coupling efficiency and outcoupling off the chip to free space. We primarily use the InP-made half of the circular Bragg grating (CBG) attached to the end of the Si WG. The CBG allows the outcoupling of QD-delivered photons travelling through the SOI chip interior. Moreover, for a selected device with a QD showing \SI{80}{\percent} single photon emission purity in the telecom C-band, we could estimate the on-chip coupling efficiency by counting the Si WG travelling photons outcoupled at the cleaved edge of the chip. In this case, the on-chip directional coupling efficiency was estimated to \SI{5.1}{\percent}. This number is less than theoretically provided, however, it is influenced by many technological factors. Expanding the proposed hybridization scheme with the method allowing for accurately positioning a QD inside the InP WG can significantly increase the on-chip coupling efficiency.

\begin{backmatter}
\bmsection{Funding}
Narodowe Centrum Nauki (2020/36/T/ST5/00511, 2020/39/D/ST5/02952); Danmarks Grundforskningsfond (DNRF147).

\bmsection{Acknowledgments}
We acknowledge financial support from the National Science Centre (Poland) within Project No. 2020/39/D/ST5/02952 and from the Danish National Research Foundation via the Research Centers of Excellence NanoPhoton (DNRF147). P. H. was supported by the Polish National Science Center within the Etiuda 8 scholarship (Grant No. 2020/36/T/ST5/00511).
\bmsection{Disclosures}
The authors declare no conflicts of interest.
\bmsection{Data Availability}
Data underlying the results presented in this paper are not publicly available at this time but may be obtained from the authors upon reasonable request.
\end{backmatter}

\bibliography{Main_text}

\end{document}


\maketitle

\section{QD growth}

The QDs are grown in the low-pressure MOVPE TurboDisc® reactor using arsine (\ce{AsH3}), phosphine (\ce{PH3}), tertiarybutylphosphine (\ce{TBP}) and trimethylindium (\ce{TMIn}) precursors with \ce{H2} as a carrier gas. The growth sequence starts with the deposition of a 0.5 $\mu$m-thick InP buffer layer on a (001)-oriented InP substrate at \SI{610}{\degreeCelsius} subsequently epitaxially covered by a 200 nm-thick In\textsubscript{0.53}Ga\textsubscript{0.47}As sacrificial layer lattice-matched to InP and a 244 nm-thick InP layer. Then, the temperature is decreased to \SI{483}{\degreeCelsius}, stabilized under TBP for 180 s and \ce{AsH3} for 27 s. Finally, nucleation of QDs occurs in the Stranski-Krastanov growth mode after deposition of nominally 0.93 mono-layer thick InAs under TMIn and \ce{AsH3} flow rates of \SI{11.8}{\micro\mole\per\minute} and \SI{52.2}{\micro\mole\per\minute}, respectively. Nucleated QDs are annealed for 60 s at the growth temperature in \ce{AsH3} ambient, before the temperature is increased to \SI{515}{\degreeCelsius} and the annealing continues for another \SI{30}{\second}. Deposition of a 244 nm-thick InP capping layer (\SI{12}{nm} at \SI{515}{\degreeCelsius}, and the remaining \SI{232}{nm} after increasing the temperature up to \SI{610}{\degreeCelsius}) finishes the growth sequence. In this procedure we achieved InAs/InP QDs of low density (\SI{3.1e8}{cm^{-2}}) characterized with high purity of single photon emission, with high extraction efficiency when integrated with photonic structures shown previously~\cite{Holewa2020}.

\section{FDTD simulations}

A primary benefit of using the FDTD approach for photonic systems containing QDs as light sources, is a broadband solution obtained in a single simulation. However, the large size of the computational domain is a drawback which extends the simulation time, so the mesh settings must be verified first, to achieve accurate results within a minute-time scale. Having a mesh cell of around 40x40x40 nm\textsuperscript{3}, we did not observe any computational staircasing errors like atypical diffraction of the field propagating along the taper at \SI{1.55}{\micro\m} wavelength. It results in 10 mesh cells per wavelength in InP. The simulation time of each taper takes 5-10 min, where we used an auto shut-off feature that stops the simulations when the energy level in the simulation region falls below 10\textsuperscript{-5}. The computational hardware is a standard personal computer with 20GB RAM and the i5-6300HQ Intel processor. Other important FDTD parameters are: default stretched coordinate PML (based on the formulation proposed by Gedney and Zhao~\cite{Gedney2010} as boundary conditions – 8 layers, simulation time 1000 fs, dt = 0.0779186 fs. In addition we must search the parameter space for the taper optimization on a stepwise approach, so we first did the quick scan of the taper geometry for a fixed length of the taper of \SI{20}{\micro\m}, then we performed a scan including additional iteration with respect to the taper length from \SI{20}{\micro\m} to \SI{25}{\micro\m}, and last we investigated the finest parameter space for the InP taper and Si waveguide width.

The dipole-WG mode coupling is examined by simulations performed for the \SI{35}{\micro\m}-long hybrid WG structure (convergence is observed around \SI{30}{\micro\m}) with emission from the in-plane polarized dipole source embedded in the center of the InP WG (x = \SI{0}{\micro\m}; z = \SI{0.51}{\micro\m}). The light field propagates over \SI{35}{\micro\m}, and is finally collected by the monitor at y = \SI{35}{\micro\m}. Then, the overlap $\eta$ between the scattered field {E,H} and the guided mode {Ef,Hf} is calculated using\cite{Davanco2009}:
\begin{equation}
    \eta=\frac{Re\left \{ \int \int \left ( \mathbf{E_f\times H^*} \right )\cdot \hat{y}dS\int \int \left ( \mathbf{E\times H_f^* }\right )\cdot \hat{y}dS \right \}}{Re\left \{ \int \int \left ( \mathbf{E_f\times H_f^*} \right )\cdot \hat{y}dS\right \}Re\left \{\int \int \left ( \mathbf{E\times H^* }\right )\cdot \hat{y}dS \right \}}
\end{equation}

The total overlap of dipole emission is calculated as a sum for of all propagating modes. As the light emission from a point source embedded in the center of the InP WG is bidirectional, the effective dipole coupling evaluated for our system is $\frac{\eta}{2}$. On the other hand, applying a grating system on one side of the WG \cite{Hepp2018} or setting a dipole in a chiral position\cite{Mrowiski2019} could significantly enhance the directionality of emission and improve the on-chip coupling of QD emission.
\begin{figure}[htbp]
\centering
\includegraphics[width=.6\linewidth]{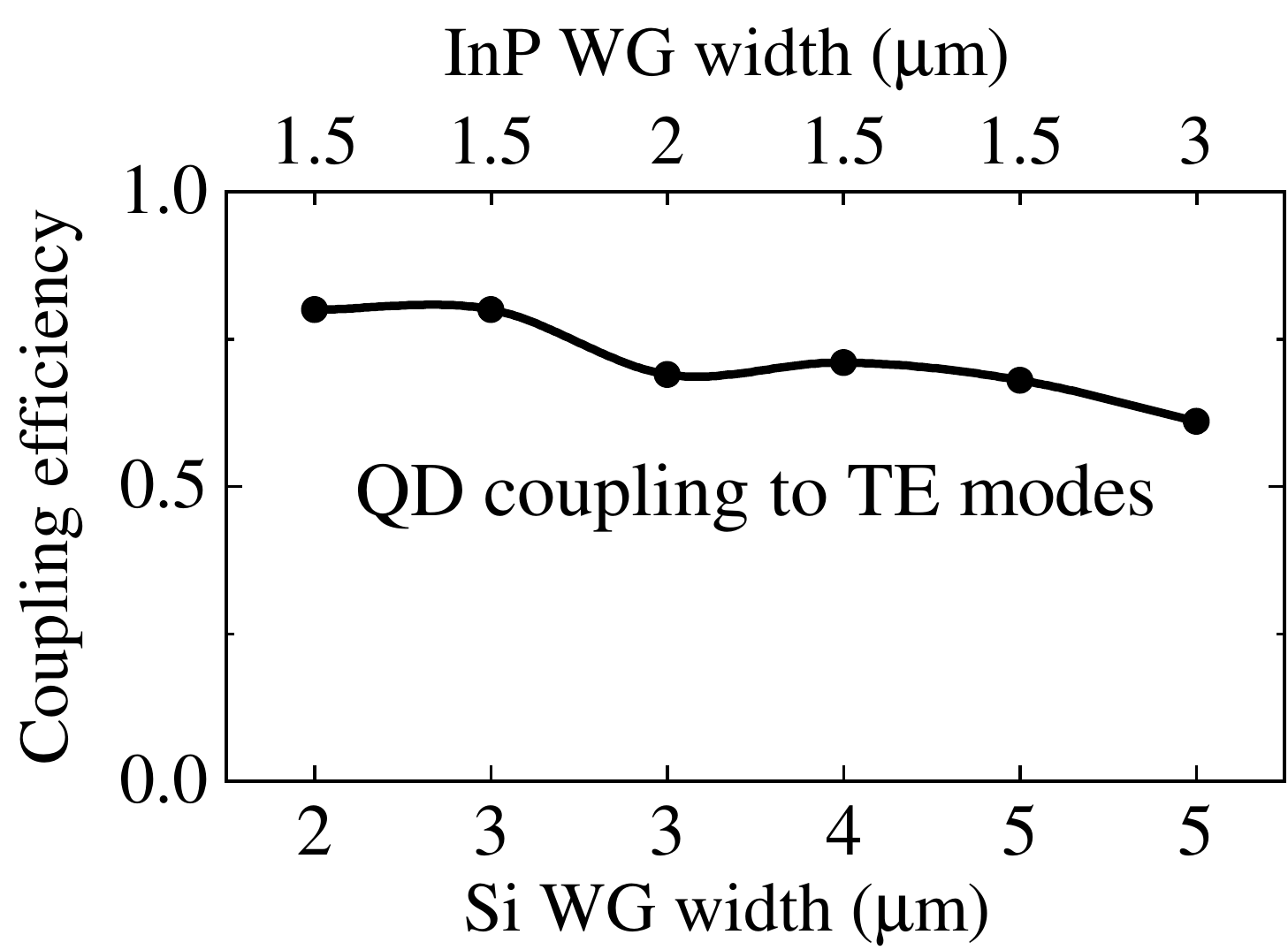}
\caption{Calculated bidirectional coupling efficiency of quantum dot emission to propagating TE-like modes in a hybrid waveguide system.}
\label{fig:S1}
\end{figure}

\section{Optical setup}
To optically characterize our device, we have measured emission properties in different configurations with detection set to either from top of the waveguide, from the outcoupler and from the Si waveguide from the cleaved edge of the sample. The sample was kept in a helium flow cryostat system at 5K. For continuous wave excitation sample was excited by a nonresonant laser ($\lambda$=787 nm). To characterize the collection efficiency of our device, QDs were excited with a pulsed femtosecond laser ($\lambda$=805 nm at 76 MHz). To collect emission from three different spots three different experimental conditions were used. To collect emission from cleaved edge of the structure we used two microscope objectives in perpendicular orientation. The emission from the Si waveguide was collected using 100x objective lens with NA = 0.65 and filtered with tunable fiber-based monochromator providing spectral and spatial resolution of \SI{100}{\micro eV} and  \SI{0.6}{\micro \m}, respectively. Emission from the normal direction was collected using 20x objective lens with NA = 0.4. Power-dependent \SI{}{\micro PL} intensity of the QD in the waveguide under pulsed excitation was measured on the InGaAs CCD and under continuous wave excitation - on SSPD. The SNSPD detection system provides a temporal resolution of 50 ps and dark count rate below 100 cps.

\begin{figure}[htbp]
\centering
\includegraphics[width=.6\linewidth]{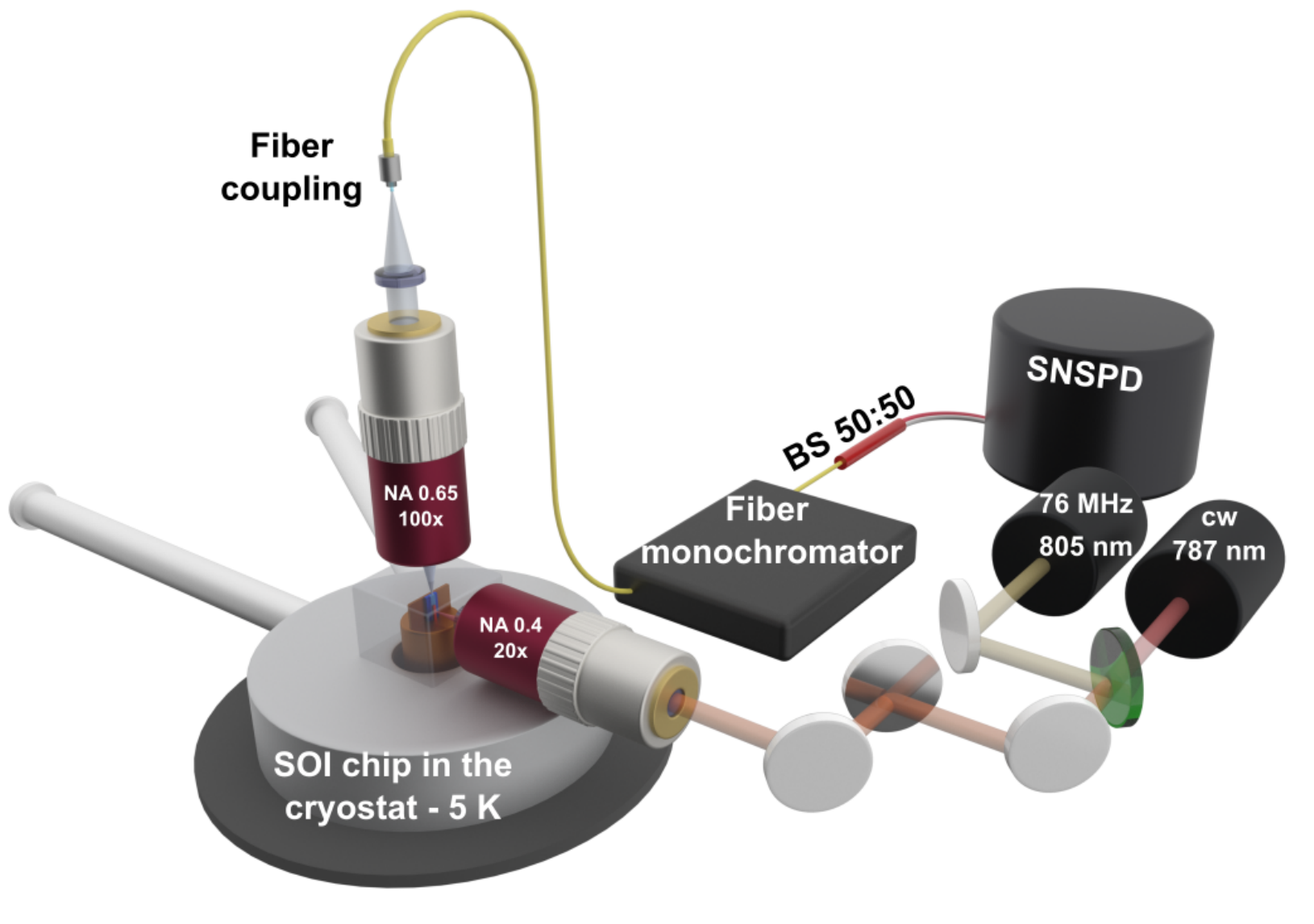}
\caption{Orthogonal arrangement of experimental setup used for evaluation of extraction efficiency from the on-chip coupled QD emission.}
\label{fig:S2}
\end{figure}

\section{Device characterization}
To determine the photon extraction efficiency from the single QD, we have measured emission rate under pulsed excitation at saturation power of the excitonic complexes and corrected it for the setup efficiency. First, we excite QD with one microscope objective (NA=0.4) from top and we detect the QD emission using another microscope objective (NA=0.4) slightly angled (adjacent to the exciting objective) and focused on the grating outcoupler. Secondly, under the same excitation conditions we detect QD emission from the QD position using the same microscope objective (NA=0.4) which is used for excitation, and third, we detect the QD emission from the cleaved facet with in-plane oriented microscope objective (NA of 0.65 for efficient coupling to SM fiber in detection path) focused on the Si WG. The extraction efficiency was measured using a free-space helium flow cryostat in the orthogonal arrangement of excitation and detection paths (see Fig. \ref{fig:S2}). The setup collection efficiency is determined using continuous-wave tuneable laser set to 1540 nm wavelength. Every free space and fiber-based optical component losses in quantum dot signal path were measured, even though the beam profiles of the laser and the QDs emission differ slightly, the objective could still gather all the photons (small divergence of the laser beam and over illumination of the objective lens by a calibration laser can lead to an overestimation of the QD extraction efficiency). We also assume only one photon is emitted per excitation pulse from excitonic complexes and ideal internal quantum efficiency which deviation from $\SI{100}{\percent}$ would only increase determined extraction efficiency. This approach enables us to assess the extraction efficiency with great accuracy. This calibration yields, a total of $\SI{0.115}{\percent}$ setup efficiency ($\SI{40}{\percent}$ objective, $\SI{6}{\percent}$ monochromator, $\SI{1}{\percent}$ multimode to single mode fiber coupling, $\SI{58}{\percent}$ rest of the optical components including efficiency of the detectors).
Low surface density of QDs makes it harder to find a QD in a waveguide of a smaller volume. Hence for experiments we comprehensively studied only multimode systems which in turn has lower coupling efficiency into single mode detection system in experimental setup due to etendue conservation principle. Etendue, denoted as $\epsilon$, can be defined as $\epsilon=S\pi NA^2$. For a waveguide with dimensions of $\numproduct{300 x 500 }\,{} \SI{ }{ \micro\m}$ and an emission angle of 90\textdegree, the etendue is calculated to be $\epsilon_{WG}=\SI{0.47}{\steradian\micro\m^{2}}$. In the experimental setup, the etendue for the single-mode fiber SMF-28 from Thorlabs is $\epsilon_{SM}=\SI{0.0103}{\steradian\micro\m^{2}}$. Consequently, the ideal coupling efficiency for such optical system can be determined to be $\frac{\epsilon_{SM}}{\epsilon_{WG}}=\SI{2.2}{\percent}$. It is important to note that this estimation serves as an upper boundary and does not account for end-facet waveguide reflections, scattering, and assumes perfect optical mode matching which for rectangular outcoupling facet does not match the Gaussian profile of optical fiber. 
To experimentally determine the coupling efficiency of the MM system with the SM system, we have coupled cw a laser at 1540 nm wavelength into the Si waveguide. The collected light from the edge of the waveguide was directed into a MM optical fiber, and the intensity was recorded using an InGaAs CCD camera. This measurement served as our reference. Subsequently, we replaced the MM fiber with a SM fiber and repeated the intensity recording. By comparing the intensity measurements with the reference measurement, we were able to estimate the coupling efficiency. Considering the reference measurement as \SI{100}{\percent}, our approach revealed a coupling efficiency of $\SI{1}{\percent}$.
Combining our etendue calculations and numerical simulations which show that the total edge outcoupling is on the order of $\SIrange{55}{60}{\percent}$ (Fig. 4c in the main text), this theoretically estimated value matches our experimentally determined value.

\begin{table}[!htb]
\centering
\caption{\bf  The transmission of the optical components in the setup used for determination of photon extraction efficiency}
\begin{tabular}{cc}
\hline
Element & Efficiency  \\
\hline
Cryostat window & \SI{90(2)}{\percent}\\
Microscope objective & \SI{40(3)}{\percent} \\
MM/SM Coupling & \SI{1.0(0.1)}{\percent} \\
Fibers and fiber connectors & \SI{61(10)}{\percent} \\
Monochromator & \SI{60(5)}{\percent}\\
SNSPD & \SI{87(3)}{\percent} \\
Total setup efficiency & \SI{0.115(0.026)}{\percent} \\
Extraction efficiency (one direction) & \SI{5.1(1.1)}{\percent} \\
Extraction efficiency (both) & \SI{10.2(2.1)}{\percent} \\
\hline
\end{tabular}
  \label{tab:Setup efficiency}
\end{table}

\bibliography{Si_main}